\documentclass[11pt,twoside]{article}

\usepackage{asp2021}

\aspSuppressVolSlug
\resetcounters

\begin{document}

\title{Astro-COLIBRI: An Innovative Platform for Real-Time Multi-Messenger Astrophysics}

\author{Bernardo Cornejo Avila,$^1$ S. Bisero,$^1$ M. Costa,$^1$ A. Ciric,$^1$ I. Jaroschewski,$^1$ W. Kiendr\'{e}b\'{e}ogo,$^1$ and F. Schussler$^1$}
\affil{$^1$IRFU, CEA, Universit\'{e} Paris-Saclay, F-91191 Gif-sur-Yvette, France; 
\email{astro.colibri@gmail.com}}

\paperauthor{Bernardo Cornejo Avila}{bernardo.cornejo@cea.fr}{0009-0003-0039-0483}{Universit\'{e} Paris-Saclay}{IRFU}{Gif-Sur-Yvette}{}{}{France}



\begin{abstract}
The discovery of transient phenomena, such as supernovae, novae, Fast Radio Bursts (FRBs), Gamma-Ray Bursts (GRBs), and stellar flares, together with the emergence of new cosmic messengers like high-energy neutrinos and Gravitational Waves (GWs), has revolutionized astrophysics in recent years. To fully exploit the scientific potential of multi-messenger and multi-wavelength follow-up observations, as well as serendipitous detections, researchers need a tool capable of rapidly compiling and contextualizing essential information for every new event. We present Astro-COLIBRI, an advanced platform designed to meet this challenge.

Astro-COLIBRI is a comprehensive platform that combines a public RESTful API, real-time databases, a cloud-based alert system, and user-friendly interfaces including a website and mobile app for iOS and Android. It ingests alerts from multiple sources in real time, applies user-defined filters, and situates each event within its multi-messenger and multi-wavelength context. The platform provides clear data visualization, concise summaries of key event properties, and evaluations of observing conditions across a wide network of observatories worldwide. We here detail the architecture of Astro-COLIBRI, from the data pipelines that manage real-time alert ingestion and processing to the design of the RESTful API, which enables seamless integration with other astronomical software and services.
\end{abstract}

\vspace{-2em}

\section{Introduction}

Time-domain astrophysics is a rapidly growing field dedicated to the study of transient phenomena. The increasing interest in well-established science cases, such as supernovae, GRBs, or active galactic nuclei (AGN) flares, together with the discovery of new phenomena such as FRBs or GWs detections, has led to a significant rise in the number of events of interest over increasingly shorter timescales. In parallel, the continuous growth of infrastructure dedicated to transient astrophysics has expanded the amount of information available for each event.

The current observational strategy is based on two types of infrastructure. \textit{All-sky} observatories continuously monitor the night sky and issue public alerts when an event is detected. These alerts are distributed through various brokers (GCN\footnote{\url{https://gcn.nasa.gov/}}, Fink\footnote{\url{https://fink-broker.org/}}, TNS\footnote{\url{https://www.wis-tns.org/}}) and received by \textit{small field-of-view} observatories, which may or may not perform follow-up observations. The decision to follow up on an alert can be automatic (e.g. \citet{hoischen_hess_2022}, \citet{collins_transients_2025}) or manual, depending on the procedures defined by each collaboration and on each different science case. This approach enables a comprehensive characterization of the source by combining multi-wavelength and multi-messenger data, but relies on efficient communication in a field where the volume of information and the number of tools and platforms are growing rapidly.

In this context, Astro-COLIBRI is a state-of-the-art tool that gathers all of the information necessary for the follow-up of a transient event in a user-friendly interface, available through both a web and mobile application. The platform continuously listens to public alert streams, filters them according to user needs, and places them into a multi-wavelength and multi-messenger context. Users are notified in real time through push notifications, enabling a fast and coordinated response to events of interest.

\section{Astro-COLIBRI}

\subsection{Back-end architecture}

Astro-COLIBRI combines a public RESTful API, real-time databases, a cloud-alert system, and user-friendly clients, including a web interface and a mobile application. Integration of these components is achieved through a modular architecture that follows a well-defined pipeline structured into successive processing steps.

\articlefigure[width=1\textwidth]{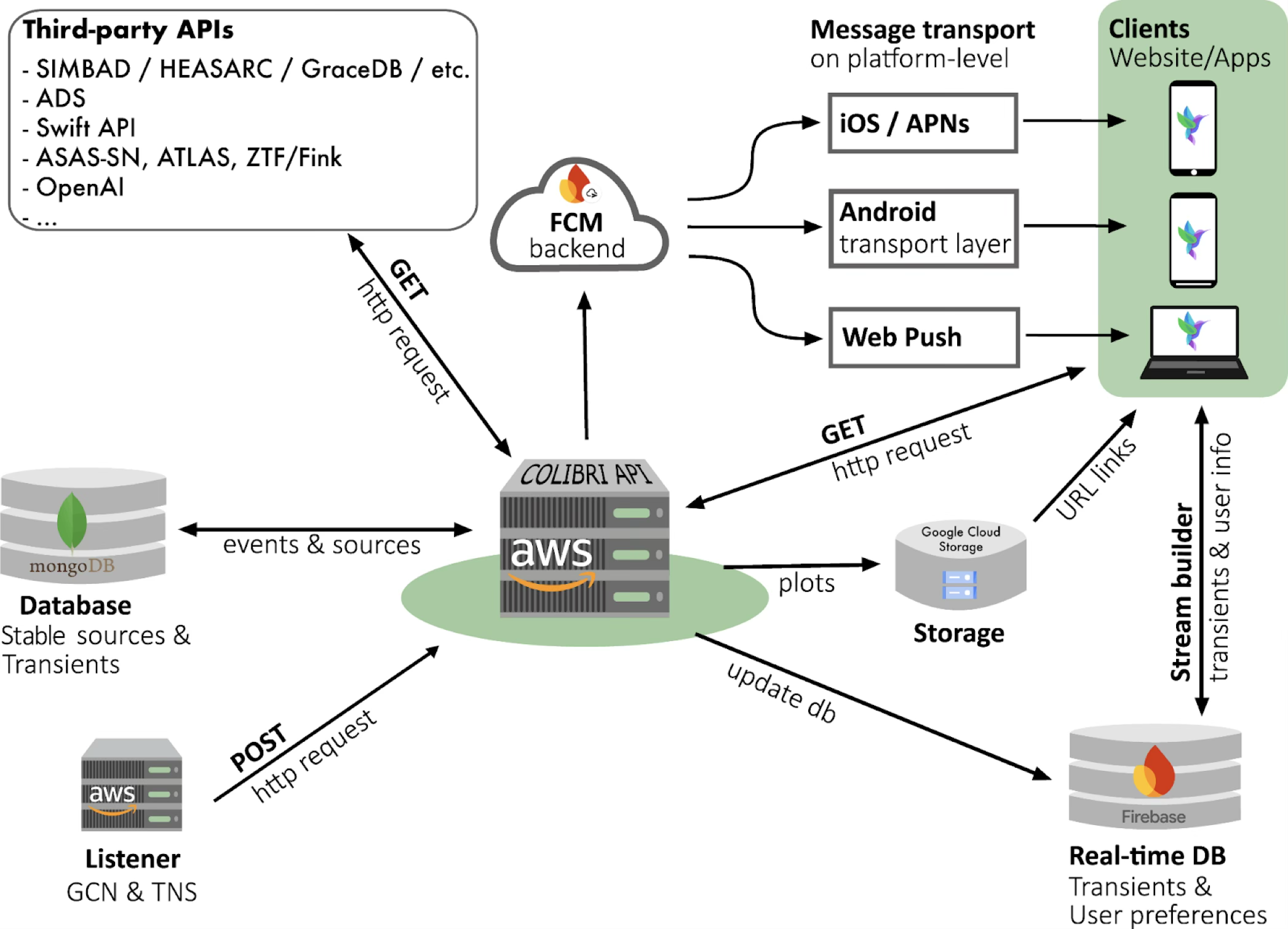}{ac_archi}{Architecture of Astro-COLIBRI illustrating the main system modules and their interactions, as described in detail in the text. Components highlighted in green are public interfaces. Inspired by \citet{reichherzer_astro-colibricoincidence_2021}.}

\begin{enumerate}
    \item \textit{RESTful API}: Fully cloud-based API developed in Python using the Flask framework and running on a dedicated AWS server. It acts as the central layer of the platform, managing internal functionalities, coordinating communication between subsystems, and providing controlled access to data. It is built around a set of endpoints that handle event processing, database interactions and user requests. The API is also responsible for event-specific plot production and the communication with external third-party services. The majority of endpoints are publicly accessible and fully documented\footnote{\url{www.astro-colibri.science/apidoc}}, with only database-modifying operations being restricted.
    \item \textit{Listener}: Continuously running on a dedicated AWS server, the event listener handles different alert streams, including those distributed by GCN, TNS, and other platforms providing VOEvent-formatted alerts. The latter are handled via the comet broker or Kafka streams, while GCN JSON notices are transmitted exclusively via Kafka. GCN circulars, TNS notifications, and email-based alerts are ingested using custom Python parsers. Incoming alerts are translated into a unified JSON-based Astro-COLIBRI format. If the trigger ID already exists in our database, the alert is deemed an update. The event is then forwarded to the API via POST requests for further processing. 
    \item \textit{External APIs}: Before being stored, events get enriched using information from external platforms. This includes links to external pages (e.g. Simbad, GraceDB), public photometric data for optical transients (e.g. from ATLAS, ZTF, ASAS-SN), and public images provided by the observatories, such as IceCube neutrino track reconstruction or Fermi-GBM light curves. Additional figures may be generated to place events into their archival context. An on-demand query is used for some external services, such as the tiling and scheduling tool for poorly localized events \texttt{tilepy} (\citet{seglar-arroyo_cross_2024}, \citet{117_adassxxxv}).
    \item \textit{Dual Database}: Astro-COLIBRI uses a dual-database architecture optimized for different requirements. Event data are stored as JSON-structured documents organized in collections, allowing flexible handling of heterogeneous and nested data structures. The primary database is hosted on MongoDB and serves as the backend for event processing, historical storage, and user queries, such as cone searches. In parallel, a real-time database hosted in Firebase provides low-latency data streams to all connected clients. It enables instantaneous propagation of new events and updates the interfaces without requiring page reloads, ensuring minimal cadence in the communication of a new alert following its publication. Both databases are continuously updated by the API.
    \item \textit{Push notifications}: User notifications are distributed through Firebase Cloud Messaging (FCM). When a new event or update fulfills the criteria of a predefined notification stream, a push notification is distributed to users subscribed to that stream.
\end{enumerate}

\subsection{Front-end}

This comprehensive structure comes together in a user-friendly front-end, built using the Flutter framework, available as a web interface (\url{https://astro-colibri.com/}) and as a mobile application for Android and iOS. These clients provide easy access to the latest events, along with filters allowing users to select only those relevant to their interests. Information about each event and its potential follow-up is condensed and easily accessible. Events are visually represented in a customizable skymap.

Main Astro-COLIBRI functionalities, like cone search, event search or visibility plots, are described in detail in \citet{reichherzer_astro-colibri_2023}. New features include a photometric summary of optical sources, additional event types along with associated filter and notification capabilities, collaboration-specific ToO submission, among many other improvements. 


\section{Outlook}

The field of multi-messenger, multi-wavelength time domain astrophysics is a constantly evolving environment. Ever-changing observation conditions, the commissioning of new facilities, and the discovery of new events all require an adaptive tool capable of responding to this rapid evolution. Built as a modular platform, Astro-COLIBRI allows for the easy implementation of updates and plug-ins, reacting promptly to emerging needs. New features and improvements are constantly being implemented to maximize the information provided to the user for each event. Finally, we are an active and accessible development team, continuously attentive to user feedback, which plays a key role in helping us build a better tool for the community.

\section*{Acknowledgments}

The authors acknowledge the support of the French Agence Nationale de la Recherche (ANR) via the MOTS project (reference ANR-22-CE31-0012). This work is also supported by the European Union's Horizon Europe Research and Innovation program under the ACME project (grant agreement n. 101131928).

\bibliography{123}  


\end{document}